%% file: 0_main.tex
\documentclass{elsarticle}
\usepackage{hyperref}
\usepackage{graphicx}  
\usepackage{dcolumn}   
\usepackage{bm}        
\usepackage{amssymb}   
\usepackage{amsfonts}
\usepackage{graphics}
\usepackage{grffile}   
\usepackage{epsfig}
\usepackage{units}
\usepackage[usenames]{color}
\usepackage[normalem]{ulem} 
\usepackage{color}
\usepackage[utf8]{inputenc}
\usepackage[T1]{fontenc}

\usepackage{subcaption}
\usepackage{siunitx}
\usepackage[dvipsnames]{xcolor}
\usepackage{tikz}
\usepackage{environ}
\makeatletter
    \newsavebox{\measure@tikzpicture}
        \NewEnviron{scaletikzpicturetowidth}[1]{%
            \def\tikz@width{#1}%
            \def\tikzscale{1}\begin{lrbox}{\measure@tikzpicture}%
            \BODY
            \end{lrbox}%
            \pgfmathparse{#1/\wd\measure@tikzpicture}%
            \edef\tikzscale{\pgfmathresult}%
            \BODY
        }
\makeatother
\usetikzlibrary{math}
\usetikzlibrary{calc}
\usetikzlibrary{decorations.pathreplacing, decorations.markings}
\usetikzlibrary{patterns, arrows, arrows.meta}

\DeclareSIUnit\ele{e^{\text{-}}}

\journal{Nucl. Instrum. Methods Phys. Res. A}
\begin{document}%
\begin{frontmatter}
\title{First demonstration of in-beam performance of bent Monolithic Active Pixel Sensors}

\author{ALICE ITS project\corref{cor1}}

\cortext[cor1]{See Appendix~\ref{sec:authors} for the complete list of authors.}

\begin{abstract}

A novel approach for designing the next generation of vertex detectors foresees to employ wafer-scale sensors that can be bent to truly cylindrical geometries after thinning them to thicknesses of \SIrange{20}{40}{\um}. To solidify this concept, the feasibility of operating bent MAPS was demonstrated using \SI{1.5x3}{\cm} ALPIDE chips. Already with their thickness of \SI{50}{\um}, they can be successfully bent to radii of about \SI{2}{\cm} without any signs of mechanical or electrical damage. During a subsequent characterisation using a \SI{5.4}{\GeV} electron beam, it was further confirmed that they preserve their full electrical functionality as well as particle detection performance.

In this article, the bending procedure and the setup used for characterisation are detailed. Furthermore, the analysis of the beam test, including the measurement of the detection efficiency as a function of beam position and local inclination angle, is discussed. The results show that the sensors maintain their excellent performance after bending to radii of \SI{2}{cm}, with detection efficiencies above \SI{99.9}{\percent} at typical operating conditions, paving the way towards a new class of detectors with unprecedented low material budget and ideal geometrical properties.
\end{abstract}

\begin{keyword}
Monolithic Active Pixel Sensors \sep Solid state detectors \sep Bent sensors
\end{keyword}

\end{frontmatter}

%
%

\input{figures/tikz_set.tex}
\input{1_intro}

\input{2_dut}

\input{3_setup}

\input{4_analysis}
\input{5_results}
\input{6_summary}

\newenvironment{acknowledgement}{\relax}{\relax}
\begin{acknowledgement}
\section*{Acknowledgements}
\input{acknowledgements}    
\end{acknowledgement}
%
\appendix
\input{authors}

\bibliography{references}

\end{document}

%% file: figures/tikz_set.tex
\tikzset{
	beamarrow/.style={
		decoration={
			markings,mark=at position 1 with 
			{\arrow[scale=2,>=stealth]{>}}
		},postaction={decorate}
	}
}
\tikzset{
	pics/.cd,
	vector out/.style={
		code={
		\draw[#1, thick] (0,0)  circle (0.15) (45:0.15) -- (225:0.15) (135:0.15) -- (315:0.15);
		}
	}
}
\tikzset{
	pics/.cd,
	vector in/.style={
		code={
		\draw[#1, thick] (0,0)  circle (0.15);
		 \fill[#1] (0,0)  circle (.05);
		 }
	}
}
\tikzset{
	global scale/.style={
		scale=#1,
		every node/.style={scale=#1}
	}
}

%% file: 1_intro.tex
\section{Introduction}
The precision of barrel vertex detectors is mainly determined by three contributions: their radial distance to the interaction point, their material budget, and their intrinsic sensor resolution. In order to achieve hermiticity, they are typically built out of detector staves placed in layers around the beam pipe. This arrangement effectively sets a practical limit on the first two factors. ALICE, for instance achieves an average radial position of~\SI{24}{\mm} and a material budget of~$\SI{0.3}{\percent}X_0$ for its new Inner Tracking System (ITS2) \cite{TDR}. More than~\SI{80}{\percent} of the material is due to the support structure, and the average distance is determined by the chip's active area in $r\varphi$-directions of \SI{1.3}{\cm} together with the need of some overlap to catch particles passing at various angles. 

A way to vastly improve these figures of merit is to use truly cylindrical detection layers made of wafer-size chips. This would not only allow placing them closer to the beam pipe, but would also largely eliminate the need for the support structure, and in turn would largely minimise the material budget to essentially that of the sensor itself. This idea is the gist of the ITS3, a proposal by the ALICE Collaboration for a novel vertex detector consisting of curved, wafer-scale, ultra-thin silicon sensors arranged in perfectly cylindrical layers, with the innermost layer positioned at a radial distance of only \SI{18}{\mm} from the nominal interaction point~\cite{loi}.

A major R\&D milestone towards these new detectors is the proof of concept of bent Monolithic Active Pixel Sensor (MAPS). Using readily available ALPIDE chips (the MAPS of the ALICE ITS2~\cite{ALPIDE-proceedings-1, ALPIDE-proceedings-2,ALPIDE-proceedings-3}, Section~\ref{sec:dut}), mechanical (Section~\ref{sec:bending}) and electrical (Section~\ref{sec:lab}) studies as well as a beam test with \SI{5.4}{\GeV} electrons (Sections~\ref{sec:setup}--\ref{sec:results}) were carried out.

%% file: 2_dut.tex
\section{The bent ALPIDE chip}
\label{sec:dut}

The ALPIDE sensor was developed by the ALICE Collaboration for its Inner Tracking System~(ITS2) \cite{ALPIDE-proceedings-1, ALPIDE-proceedings-2,ALPIDE-proceedings-3}. The chip is produced in the \SI{180}{\nm}~CMOS process of TowerJazz~\cite{towerjazz} featuring a \SI{25}{\um}-thick epitaxial layer. Here, chips thinned to \SI{50}{\um} have been used.

\begin{figure}[thp]
	\centering
    \includegraphics[width=\textwidth]{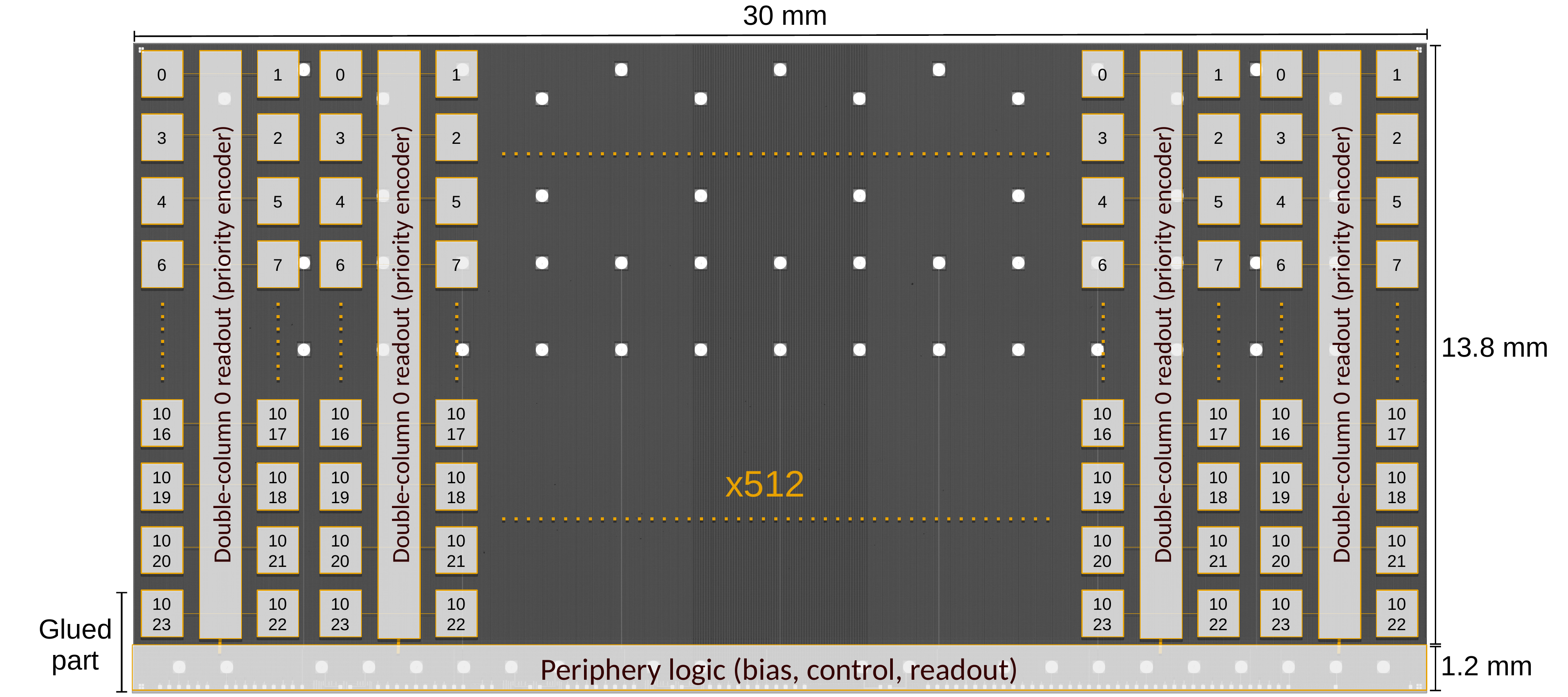}
	\caption{Layout of the ALPIDE pixel matrix. The pixels are organised in double-columns, each featuring a priority encoder circuit which propagates the addresses of the hit pixels to the periphery logic. The aluminum pads providing the electrical interface to the chip are located on the top of the periphery logic.}
	\label{fig:alpide}
\end{figure}

ALPIDE features a matrix of \num{1024x512} (column~$\times$~row) pixels with binary output. The pixels of size \SI{26.88x29.24}{\um} are organised in double-columns, each one having \num{1024}~pixels (Fig.~\ref{fig:alpide}). The central part of each double-column is occupied by priority encoding circuits which propagate the addresses of the hit pixels to the digital circuity on the chip periphery. The digital periphery occupies an area of \SI{1.2x30}{\mm} along the edge of the chip.
A series of aluminum pads on top of the digital periphery, close to the edge of the chip, provide the electrical interface to the chip. 

Each ALPIDE pixel contains the sensing diode connected to its individual and continuously active front-end amplifier, shaper, discriminator, and multiple-event buffer. It also contains analog and digital testing circuitry, allowing the measurement of the charge threshold by injecting a programmable test charge into the sensing node. The threshold can be changed for all pixels simultaneously by adjusting the amplifier parameters~\cite{ALPIDE-proceedings-1, ALPIDE-proceedings-2, ALPIDE-proceedings-3, pulselength}. 

At this point, it is worth pointing out that the pixel matrix itself -- which will be the part of the chip that is bent, see below -- is a highly integrated circuit, including analog and digital circuitry totaling to order of~\num{200}~transistors per pixel cell and with a dense metal routing.

\subsection{Bending procedure} 
\label{sec:bending}

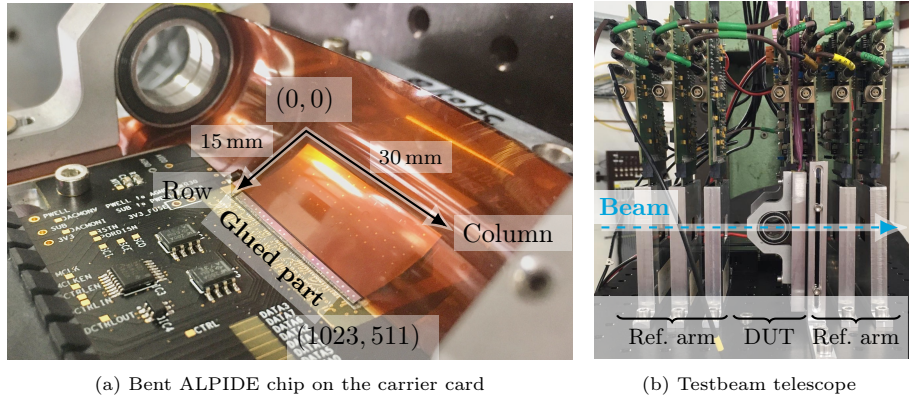
\begin{figure}[thp]
	\centering
	\begin{subfigure}[t]{0.62\textwidth}
	    \input{figures/bent_alpide_image.tex}
	    \caption{Bent ALPIDE chip on the carrier card}
	    \label{fig:bent_alpide_carrier}
	\end{subfigure}
	\begin{subfigure}[t]{0.368\textwidth}
	    \input{figures/telescope_side_image.tex}
	    \caption{Testbeam telescope}
	    \label{fig:setup_side}
	\end{subfigure}
	\caption{The ALPIDE chip glued to the carrier card and held in bent position via two polyamide foils attached to the aluminum wheels (a) and the same arrangement inserted in the testbeam telescope consisting of six flat ALPIDE tracking planes (b).}
	\label{fig:bent_alpide}
\end{figure}

A procedure to bend the chip in a progressive and reversible manner was specifically developed and is described as follows. 
The long edge hosting the bonding pads and the periphery logic and about \SI{0.8}{\mm}-wide strip of the pixel matrix (Fig.~\ref{fig:alpide}) is glued onto a carrier board for a \SI{2}{\mm}-wide section (Fig.~\ref{fig:bent_alpide_carrier}) by means of acrylic adhesive\footnote{3M\texttrademark Adhesive Transfer Tape 467MP, \href{https://www.3m.com/3M/en_US/company-us/all-3m-products/~/3M-Adhesive-Transfer-Tape-467MP?N=5002385+3293242532&rt=rud}{https://www.3m.com/}}. The chip is then wire-bonded to the carrier card, before executing the bending procedure. The bonding area remains flat and well secured throughout and after the procedure. The rest of the chip is left unattached and is lightly compressed between two layers of \SI{120}{\um}-thick polyimide foil (Fig.~\ref{fig:bent_alpide_carrier}).
The polyimide foil layers are attached to two lateral wheels (Fig.\ref{fig:bent_alpide_carrier}) that can be moved in parallel to the short edge of the chip by means of a micrometer-precision positioning system (not visible in pictures). By moving the wheels towards the chip, the polyimide foils wrap around them, and bend the chip into a cylindrical shape.
Once the desired curvature is achieved, the wheels position is fixed using the $\Omega$-shaped aluminum fixtures (visible in Fig.~\ref{fig:bent_alpide_carrier}). Alternatively, the micro-positioning system allows reverting to the flat or any intermediate position.

\subsection{Curvature measurement}
\label{sec:curvature_measurement}

\begin{figure}[thp]
	\centering
    \includegraphics[width=0.7\textwidth]{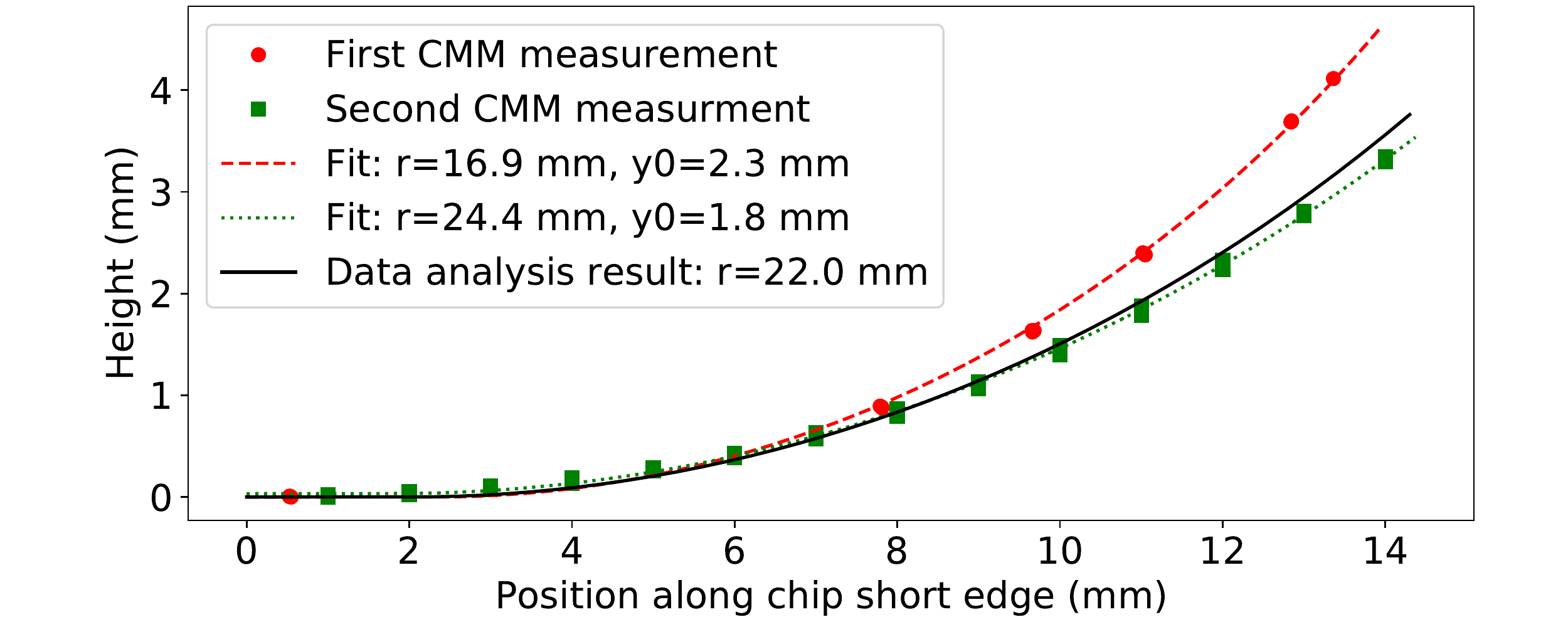}
	\caption{Curvature radius measured using a Coordinate Measuring Machine (CMM) and obtained from the testbeam data analysis (Section~\ref{sec:single-run-analysis}).}
	\label{fig:radius}
\end{figure}

A 3-D metrological mapping of the chip surface was performed using a Coordinate Measuring Machine (CMM) before and after the testbeam measurement.
The carrier card with the bent chip (Fig.\ref{fig:bent_alpide_carrier}) was laid on the measurement table and a series of data points was measured with an optical head, providing a resolution of \SI{5}{\um} in the table plane, and \SI{80}{\um} on the height.
The data point series was projected on the axis given by the short edge of the chip, as reported in Fig.~\ref{fig:radius}, and fitted with a circle of radius $r$ and origin in $y_0$, with $y_0$ parameter describing the flat (glued) part of the chip.
The fit procedure provided an average curvature radius of \SI{16.9}{\mm} before and \SI{24.4}{\mm} after the testbeam.
The change in curvature is attributed to a relaxation of the polyimide foil holder before, during, or after the testbeam. Therefore, it was expected, and later confirmed by the data analysis (Section~\ref{sec:single-run-analysis}), that the curvature during the testbeam measurement was in-between the two CMM measurements.

\subsection{Performance comparison before and after bending}
\label{sec:lab}

\begin{figure}[thp]
    \centering
    \includegraphics[width=0.7\textwidth]{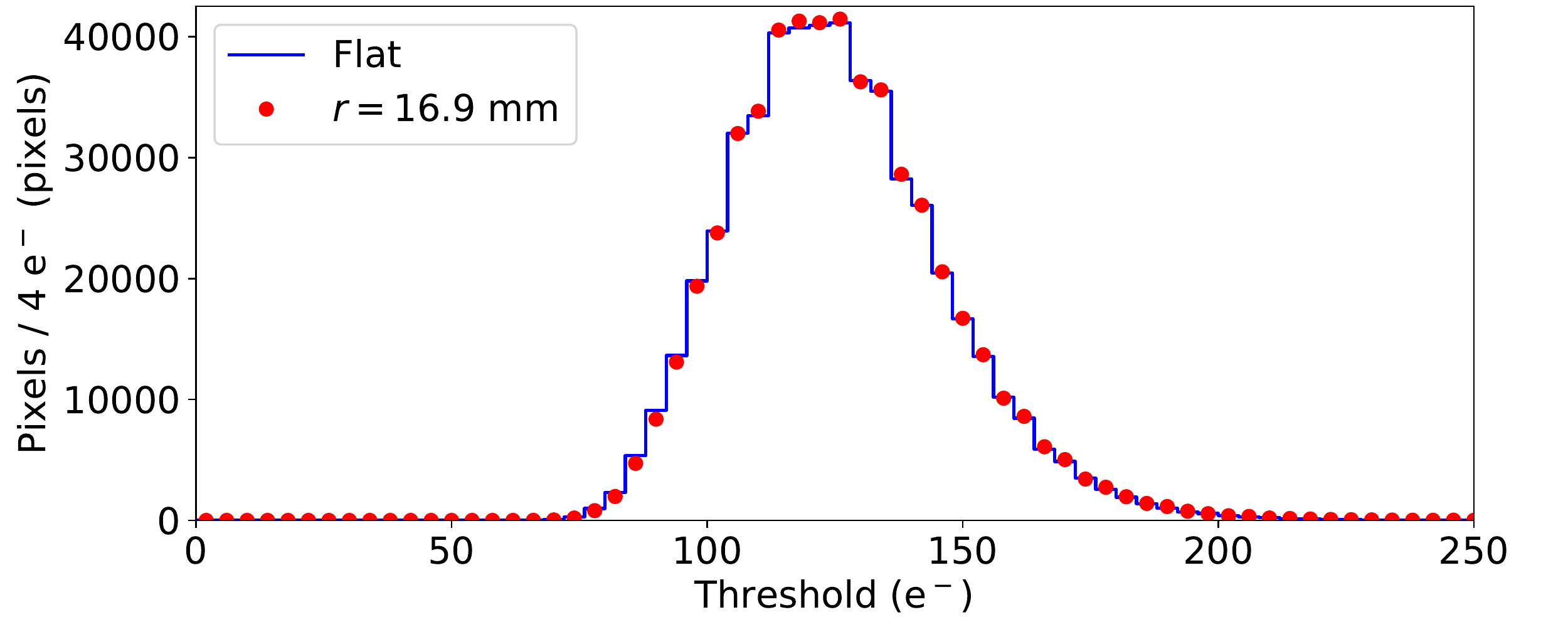}
	\caption{Pixel threshold distribution before and after bending the chip. The difference between the two measurements is negligible.}
	\label{fig:threshold-distribution}
\end{figure}

To verify the electrical functionality, i.e.~the analogue in-pixel circuitry and the digital column circuitry propagating the position of the hit pixels to the periphery, the chip was characterised in terms of number of non-responsive pixels, pixel thresholds, noise and fake-hit rate before and after the bending. 
The tested parameters are unchanged or their change is negligible, as is shown in Fig.~\ref{fig:threshold-distribution} for the pixel threshold distribution, as an example.

Since this measurement exercises the full analog and digital processing chain of the chip and hence it does not only show that its front-end performance is unaffected by the bending, but also that the whole circuit -- notably the in-matrix distributed digital readout network -- is still in full function. 

%% file: figures/bent_alpide_image.tex
\begin{scaletikzpicturetowidth}{\textwidth}
	\begin{tikzpicture}[decoration=brace, global scale = \tikzscale]
		\node[anchor=south west,inner sep=0] at (0,0) {
			\includegraphics[width=\textwidth]{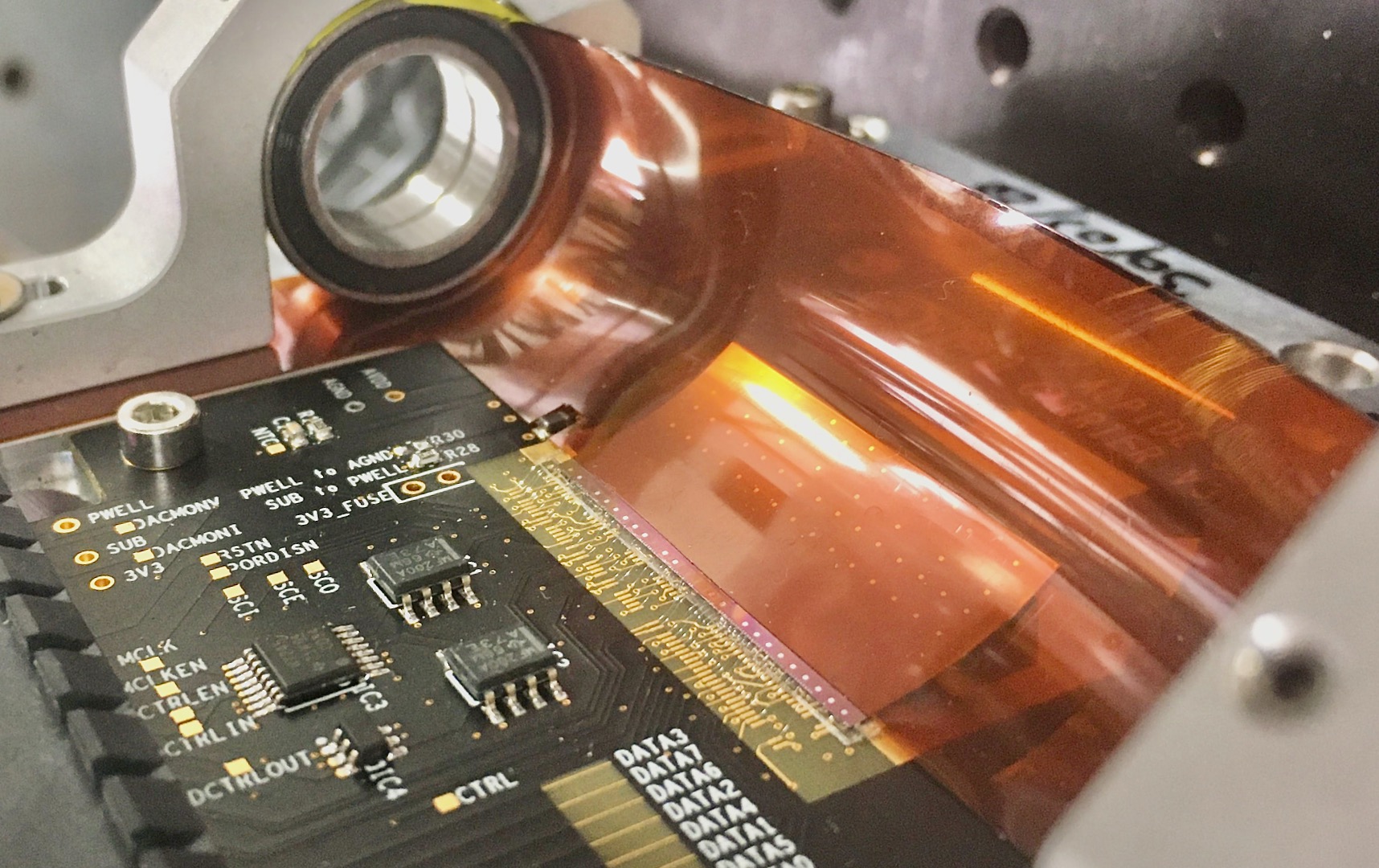}
		};
		
		\tikzmath{
			\ArrWidth = 1.75mm;
			\ArrLength = \ArrWidth * 1.2;
			\Opacity=0.4;
			\Opacitytwo=0.5;
			\ArrWidthtwo = \ArrWidth * 2;
			\ArrLengthtwo = \ArrLength * 2;
		}
	
		\coordinate (T) at (0.535\textwidth,0.39\textwidth);
		\coordinate (L) at (0.417\textwidth,0.3\textwidth);
		\coordinate (R) at (0.77\textwidth,0.225\textwidth);
		\coordinate (B) at (0.63\textwidth,0.11\textwidth);

		
			
		\node[left, align=center, xshift=-0.02\textwidth, opacity=\Opacity, text opacity=1.0, fill=White] at (L) {
			 \textcolor{black}{Row}
			};
		\node[right, align=center, xshift=0.02\textwidth, opacity=\Opacity, text opacity=1.0, fill=White] at (R) {
			 \textcolor{black}{Column}
			};
		\node[above, align=center, yshift=0.03\textwidth, opacity=\Opacity, text opacity=1.0, fill=White] at (T) {
			 \textcolor{black}{$(0,0)$}
			};
		\node[below, align=center, yshift=-0.03\textwidth, opacity=\Opacity, text opacity=1.0, fill=White] at (B) {
			 \textcolor{black}{$(1023,511)$}
			};
			
		\draw[thick, fill=White, opacity=0.3] (L) -- ($(L)+(-0.13,-0.1)$) -- ($(B)+(-0.13,-0.1)$)
		node[below, xshift=-0.02\textwidth, opacity=\Opacity, text opacity=1.0, fill=White, midway, sloped] (TextNode) { \small \textcolor{black}{\textbf{Glued part}}}
		-- (B) -- cycle;
			
		\draw[{Latex[width=\ArrWidthtwo, length=\ArrLengthtwo]}-{Latex[width=\ArrWidthtwo, length=\ArrLengthtwo]}, color=White, line width=1mm, opacity=\Opacitytwo]
		     ($(L) + (-1* 0.023\textwidth,-1 * 0.01\textwidth)$)
		-- ($(T) + (-1 * 0.005\textwidth,0.015\textwidth)$)
		-- ($(R) + (0.023\textwidth,0.001\textwidth)$);
		
		\draw[{Latex[width=\ArrWidth, length=\ArrLength]}-{Latex[width=\ArrWidth, length=\ArrLength]}, color=black, thick] ($(L) + (-1* 0.01\textwidth,0)$)
			-- node[above, xshift=-0.07\textwidth,yshift=0.005\textwidth, opacity=\Opacity, text opacity=1.0, fill=White] {
				\footnotesize \textcolor{black}{\SI{15}{\milli\metre}}
				}($(T) + (-1 * 0.005\textwidth,0.015\textwidth)$)
			-- node[above,xshift=0.06\textwidth, yshift=0.01\textwidth, opacity=\Opacity, text opacity=1.0, fill=White] {
				 \footnotesize \textcolor{black}{\SI{30}{\milli\metre}}
				} ($(R) + (0.01\textwidth,0.01\textwidth)$);
	\end{tikzpicture}
\end{scaletikzpicturetowidth}

%% file: figures/telescope_side_image.tex
\begin{scaletikzpicturetowidth}{\textwidth}
	\begin{tikzpicture}[decoration=brace, global scale = \tikzscale]
		\node[anchor=south west,inner sep=0] at (0,0) {
			\scalebox{-1}[1] {
				\includegraphics[trim = {17cm 7.5cm 11cm 2.5cm}, clip, width=\textwidth]{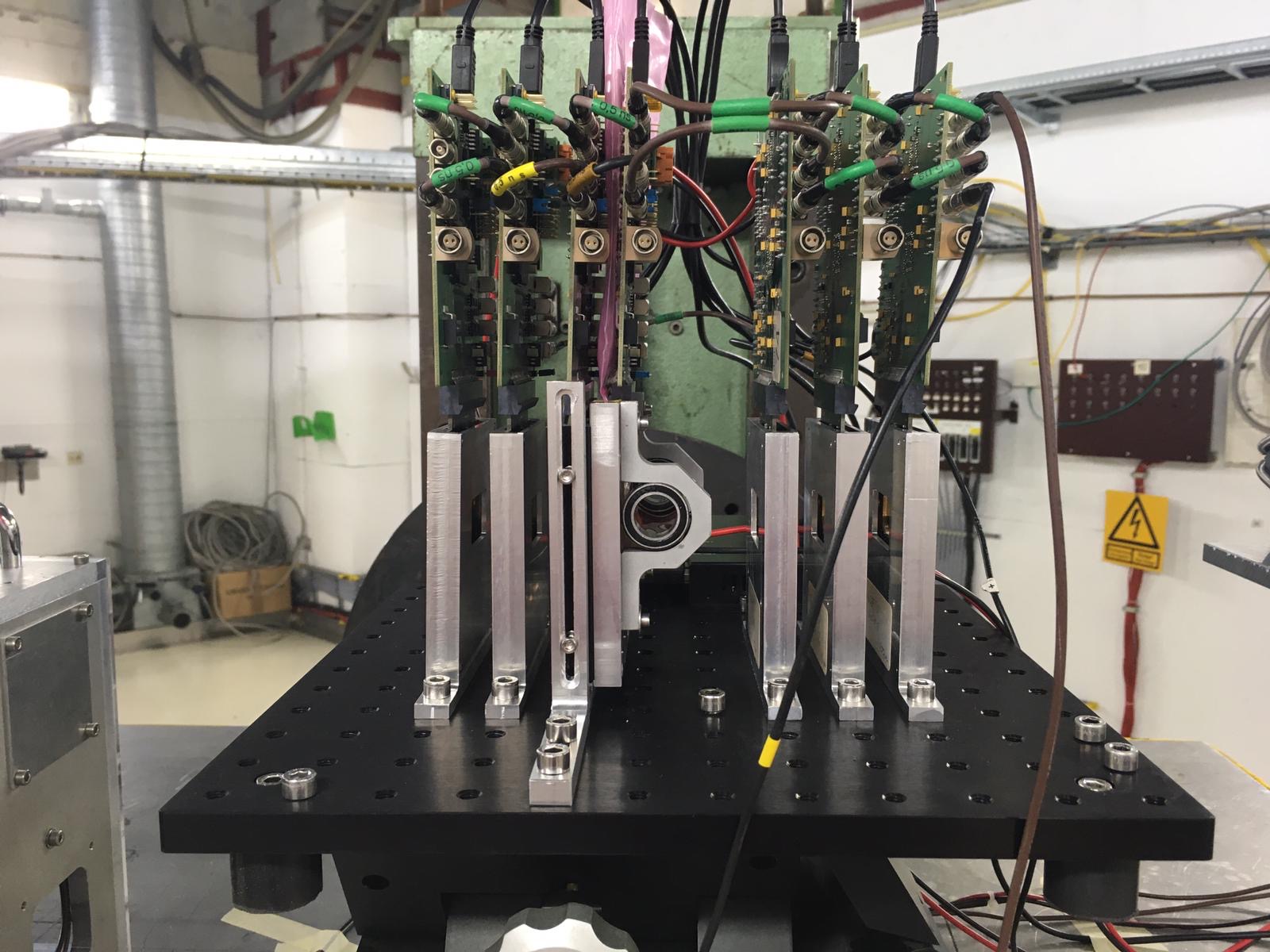}
			}
		};
		
		\tikzmath{
			\ArrWidth = 2mm;
			\ArrLength = \ArrWidth * 1.2;
			\ArrWidthtwo = \ArrWidth * 2;
			\ArrLengthtwo = \ArrLength * 2;
			\Opacity=0.7;
			\Opacitytwo=0.5;
			\Opacitythree=0.6;
			\BraceY=0.15;
			\BoxY=0.2;
		}
	
		\coordinate (start) at (0.05\textwidth,0.42\textwidth);
		\coordinate (stop) at (0.995\textwidth,0.42\textwidth);
		
		\draw[-{Latex[width=\ArrWidthtwo, length=\ArrLengthtwo]}, color=White, line width=1mm, opacity=\Opacitytwo] (start)
			-- ($(stop) + (0.04\textwidth,0)$);
		\draw[-{Latex[width=\ArrWidth, length=\ArrLength]}, color=Cyan!90!Black, thick, dashed] (start)
			-- ($(stop) - (0.35\textwidth, 0)$)
			--node[above,xshift=-0.68\textwidth, yshift=0.01\textwidth,  opacity=\Opacity, text opacity=1.0, fill=White] {
				\textbf{  \textcolor{Cyan!90!Black}{Beam}}
				} (stop);
				
		\draw[color=White, opacity=\Opacitythree, fill=White, line width=0mm] (0.018\textwidth,0)
			-- (1.025\textwidth, 0)
			-- (1.025\textwidth, \BoxY\textwidth)
			-- (0.018\textwidth, \BoxY\textwidth)
			-- cycle;
			
		\draw[decorate, yshift=-0.01\textwidth, thick]  (0.45\textwidth,\BraceY\textwidth)
			-- node[below=0.4ex, xshift=-0.00\textwidth] {\small Ref.~arm} (0.1\textwidth,\BraceY\textwidth);
		\draw[decorate, yshift=-0.01\textwidth, thick]  (\textwidth,\BraceY\textwidth)
			-- node[below=0.4ex, xshift=-0.00\textwidth] {\small Ref.~arm} (0.72\textwidth,\BraceY\textwidth);
		\draw[decorate, yshift=-0.01\textwidth, thick]  (0.7\textwidth,\BraceY\textwidth)
			-- node[below=0.4ex, xshift=-0.00\textwidth] {\small DUT} (0.47\textwidth,\BraceY\textwidth);
	\end{tikzpicture}
\end{scaletikzpicturetowidth}

%% file: 3_setup.tex
\section{Testbeam set-up}
\label{sec:setup}

\begin{figure}[thp]
	\centering
    \input{figures/setup_sketch.tex}
    \caption{Sketch of the beam telescope with the bent DUT sandwiched between six flat ALPIDE reference planes. The DUT position can be translated in the y-direction.}
    \label{fig:setup_sketch}
\end{figure}
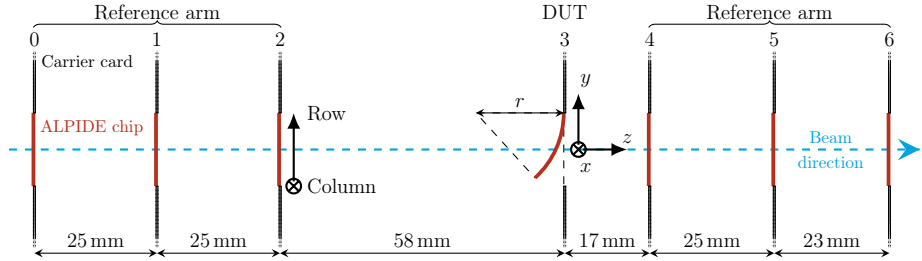

The testbeam was carried out at DESY testbeam facility beam line~24 \cite{desyII}, with a \SI{5.4}{\giga\electronvolt} electron beam.
The testbeam energy and particle species is selectable with a dipole magnet and \SI{5.4}{\giga\electronvolt} electrons were chosen.
The beam telescope comprised of \num{6}~reference planes with flat ALPIDE chips. The bent device under test (DUT) was placed in the middle, with \num{3}~reference planes on each side (Fig.~\ref{fig:setup_side} and Fig.~\ref{fig:setup_sketch}). The vertical~($y$) position of the DUT was adjustable, allowing the fine tunining of the position of the DUT with respect to the beam.

The trigger was given by the coincidence of the discriminated output of two photomultiplier tubes (operated in the plateau region) connected to two scintillators with size of \SI{4x5}{\cm}, placed in front and behind the telescope. 
The triggering logic includes an event separation time of \SI{100}{\micro\second} and a past protection time of \SI{50}{\micro\second}, i.e.~a veto on triggering in \SI{50}{\micro\second} following a scintillator output signal. The latter is added to avoid the pile-up as the ALPIDE in-pixel amplifier pulse can reach lengths of few~tens~of~\si{\us} for very low threshold values \cite{pulselength}.

The data acquisition was based on the EUDAQ framework~\cite{eudaq}. A total of 176~runs with at least 300k~events each were taken corresponding to different thresholds and DUT positions.

%% file: figures/setup_sketch.tex
\begin{scaletikzpicturetowidth}{\textwidth}
    \begin{tikzpicture}[decoration=brace, global scale = \tikzscale]
			\tikzmath{
				\BoardHeight = 1.25;
				\BoardHeightmod = \BoardHeight * 0.85;
				\BoardWidth = 0.04;
				\DutHeight = 1.5;
				\ChipWidth = 0.5mm;
				\DutRadius = 1.8;
				\DutAngle = \DutHeight / \DutRadius * 180 / 3.141592;
				\CoordinateDist = 0.3;
				\VecRad = 0.15;
				\ArrWidth = 1.75mm;
				\ArrLength = \ArrWidth * 1.2;
				\CoordinateLength = 1;
				%
				\z01 = 2.5;
				\z12 = 2.5;
				\z23 = 5.8;
				\z34 = 1.7;
				\z45 = 2.5;
				\z56 =  2.3;
				\px0 = 0;
				\pylow = \DutHeight / -2;
				\pyup = \DutHeight / 2;
				\px1 = \px0 + \z01;
				\px2 = \px1 + \z12;
				\px3 = \px2 + \z23;
				\px4 = \px3 + \z34;
				\px5 = \px4 + \z45;
				\px6 = \px5 + \z56;
			}
		
			\coordinate (p0lowUL) at (\px0,\pylow);
			\coordinate (p0upBL) at (\px0,\pyup);
			\coordinate (p1lowUL) at (\px1,\pylow);
			\coordinate (p1upBL) at (\px1,\pyup);
			\coordinate (p2lowUL) at (\px2,\pylow);
			\coordinate (p2upBL) at (\px2,\pyup);
			\coordinate (p3lowUL) at (\px3,\pylow);
			\coordinate (p3upBL) at (\px3,\pyup);
			\coordinate (p4lowUL) at (\px4,\pylow);
			\coordinate (p4upBL) at (\px4,\pyup);
			\coordinate (p5lowUL) at (\px5,\pylow);
			\coordinate (p5upBL) at (\px5,\pyup);
			\coordinate (p6lowUL) at (\px6,\pylow);
			\coordinate (p6upBL) at (\px6,\pyup);
			
			\draw[beamarrow, color=Cyan!90!Black, thick, dashed] (-0.5,0)
				--  ($(\px2 + \BoardWidth,0) $)
				-- ($(\px3,0) $)
				-- ($(\px4,0) $)
				--  ($(\px5,0) $)
				--  node[below] {\textcolor{Cyan!90!Black}{direction}}  node[above] {\textcolor{Cyan!90!Black}{Beam}}($(\px6,0) $)
				-- ($(\px6 + \BoardWidth + 0.6,0) $);
			
			\draw[line width = \ChipWidth, color=BrickRed] ($(p0upBL) + (0,0)$)
			    -- ($(p0upBL) + (0,-1 * 0.2*\DutHeight)$) node[right] {ALPIDE chip}
				-- ($(p0upBL) + (0,-1 * \DutHeight)$);
			\draw[line width = \ChipWidth, color=BrickRed] ($(p1upBL) + (0,0)$)
				-- ($(p1upBL) + (0,-1 * \DutHeight)$);
			\draw[line width = \ChipWidth, color=BrickRed] ($(p2upBL) + (0,0)$)
				-- ($(p2upBL) + (0,-1 * \DutHeight)$);
			\draw[line width = \ChipWidth, color=BrickRed] ($(p4upBL) + (0,0)$)
				-- ($(p4upBL) + (0,-1 * \DutHeight)$);
			\draw[line width = \ChipWidth, color=BrickRed] ($(p5upBL) + (0,0)$)
				-- ($(p5upBL) + (0,-1 * \DutHeight)$);
			\draw[line width = \ChipWidth, color=BrickRed] ($(p6upBL) + (0,0)$)
				-- ($(p6upBL) + (0,-1 * \DutHeight)$);

			\draw[dashed] ($(p3upBL) + (0,0)$)
				-- ($(p3upBL) + (0,-1 * \DutHeight)$);
			\draw[dashed] ($(p3upBL) + (0,0)$)
				-- ++(0+180:\DutRadius) -- +(-1 * \DutAngle:\DutRadius);
			\draw[line width = \ChipWidth, color=BrickRed] ($(p3upBL) + (0,0)$)
				arc (0:-1 * \DutAngle:\DutRadius);
			
			\fill[top color = gray!100,bottom color = gray!25, middle color=gray] (p0lowUL)
				-- ($(p0lowUL) + (\BoardWidth,0)$)
				--  ($(p0lowUL) + (\BoardWidth,-1 * \BoardHeight)$)
				-- ($(p0lowUL) + (0,-1 * \BoardHeight)$)
				-- cycle; 
			\draw [] (p0lowUL)
				edge [densely dotted]  ($(p0lowUL) + (0,-1 * \BoardHeight)$); 
			\draw [] ($(p0lowUL) + (\BoardWidth,0)$)
				edge [densely dotted]  ($(p0lowUL) + (\BoardWidth,-1 * \BoardHeight)$); 
			\draw [] ($(p0lowUL) + (0,-1 * \BoardHeightmod)$)
				-- (p0lowUL)
				-- ($(p0lowUL) + (\BoardWidth,0)$)
				-- ($(p0lowUL) + (\BoardWidth,-1 * \BoardHeightmod)$); 
				
			\fill[bottom color = gray!100,top color = gray!25, middle color=gray] (p0upBL)
				-- ($(p0upBL) + (\BoardWidth,0)$)
				--  ($(p0upBL) + (\BoardWidth,\BoardHeight)$)
				-- ($(p0upBL) + (0,\BoardHeight)$)
				-- cycle; 
			\draw [] (p0upBL)
				edge [densely dotted]  ($(p0upBL) + (0,\BoardHeight)$); 
			\draw [] ($(p0upBL) + (\BoardWidth,0)$)
				edge [densely dotted]  ($(p0upBL) + (\BoardWidth,\BoardHeight)$); 
			\draw [] ($(p0upBL) + (0,\BoardHeightmod)$)
				-- (p0upBL)
				-- ($(p0upBL) + (\BoardWidth,0)$)
				-- ($(p0upBL) + (\BoardWidth, \BoardHeightmod)$) node[right] {Carrier card}; 

			\fill[top color = gray!100,bottom color = gray!25, middle color=gray] (p1lowUL)
				-- ($(p1lowUL) + (\BoardWidth,0)$)
				--  ($(p1lowUL) + (\BoardWidth,-1 * \BoardHeight)$)
				-- ($(p1lowUL) + (0,-1 * \BoardHeight)$) 
				-- cycle; 
			\draw [] (p1lowUL)
				edge [densely dotted]  ($(p1lowUL) + (0,-1 * \BoardHeight)$); 
			\draw [] ($(p1lowUL) + (\BoardWidth,0)$)
				edge [densely dotted]  ($(p1lowUL) + (\BoardWidth,-1 * \BoardHeight)$); 
			\draw [] ($(p1lowUL) + (0,-1 * \BoardHeightmod)$)
				-- (p1lowUL)
				-- ($(p1lowUL) + (\BoardWidth,0)$)
				-- ($(p1lowUL) + (\BoardWidth,-1 * \BoardHeightmod)$); 

			\fill[bottom color = gray!100,top color = gray!25, middle color=gray] (p1upBL)
				-- ($(p1upBL) + (\BoardWidth,0)$)
				--  ($(p1upBL) + (\BoardWidth,\BoardHeight)$)
				-- ($(p1upBL) + (0,\BoardHeight)$)
				-- cycle; 
			\draw [] (p1upBL)
				edge [densely dotted]  ($(p1upBL) + (0,\BoardHeight)$); 
			\draw [] ($(p1upBL) + (\BoardWidth,0)$)
				edge [densely dotted]  ($(p1upBL) + (\BoardWidth,\BoardHeight)$); 
			\draw [] ($(p1upBL) + (0,\BoardHeightmod)$)
				-- (p1upBL)
				-- ($(p1upBL) + (\BoardWidth,0)$)
				-- ($(p1upBL) + (\BoardWidth, \BoardHeightmod)$); 

			\fill[top color = gray!100,bottom color = gray!25, middle color=gray] (p2lowUL)
				-- ($(p2lowUL) + (\BoardWidth,0)$)
				--  ($(p2lowUL) + (\BoardWidth,-1 * \BoardHeight)$)
				-- ($(p2lowUL) + (0,-1 * \BoardHeight)$)
				-- cycle; 
			\draw [] (p2lowUL)
				edge [densely dotted]  ($(p2lowUL) + (0,-1 * \BoardHeight)$); 
			\draw [] ($(p2lowUL) + (\BoardWidth,0)$)
				edge [densely dotted]  ($(p2lowUL) + (\BoardWidth,-1 * \BoardHeight)$); 
			\draw [] ($(p2lowUL) + (0,-1 * \BoardHeightmod)$)
				-- (p2lowUL)
				-- ($(p2lowUL) + (\BoardWidth,0)$)
				-- ($(p2lowUL) + (\BoardWidth,-1 * \BoardHeightmod)$); 

			\fill[bottom color = gray!100,top color = gray!25, middle color=gray] (p2upBL)
				-- ($(p2upBL) + (\BoardWidth,0)$)
				--  ($(p2upBL) + (\BoardWidth,\BoardHeight)$)
				-- ($(p2upBL) + (0,\BoardHeight)$)
				-- cycle; 
			\draw [] (p2upBL)
				edge [densely dotted]  ($(p2upBL) + (0,\BoardHeight)$); 
			\draw [] ($(p2upBL) + (\BoardWidth,0)$)
				edge [densely dotted]  ($(p2upBL) + (\BoardWidth,\BoardHeight)$); 
			\draw [] ($(p2upBL) + (0,\BoardHeightmod)$)
				-- (p2upBL)
				-- ($(p2upBL) + (\BoardWidth,0)$)
				-- ($(p2upBL) + (\BoardWidth, \BoardHeightmod)$); 

			\fill[top color = gray!100,bottom color = gray!25, middle color=gray] (p3lowUL)
				-- ($(p3lowUL) + (\BoardWidth,0)$)
				--  ($(p3lowUL) + (\BoardWidth,-1 * \BoardHeight)$)
				-- ($(p3lowUL) + (0,-1 * \BoardHeight)$)
				-- cycle; 
			\draw [] (p3lowUL)
				edge [densely dotted]  ($(p3lowUL) + (0,-1 * \BoardHeight)$); 
			\draw [] ($(p3lowUL) + (\BoardWidth,0)$)
				edge [densely dotted]  ($(p3lowUL) + (\BoardWidth,-1 * \BoardHeight)$); 
			\draw [] ($(p3lowUL) + (0,-1 * \BoardHeightmod)$)
				-- (p3lowUL)
				-- ($(p3lowUL) + (\BoardWidth,0)$)
				-- ($(p3lowUL) + (\BoardWidth,-1 * \BoardHeightmod)$); 

			\fill[bottom color = gray!100,top color = gray!25, middle color=gray] (p3upBL)
				-- ($(p3upBL) + (\BoardWidth,0)$)
				--  ($(p3upBL) + (\BoardWidth,\BoardHeight)$)
				-- ($(p3upBL) + (0,\BoardHeight)$)
				-- cycle; 
			\draw [] (p3upBL)
				edge [densely dotted]  ($(p3upBL) + (0,\BoardHeight)$); 
			\draw [] ($(p3upBL) + (\BoardWidth,0)$)
				edge [densely dotted]  ($(p3upBL) + (\BoardWidth,\BoardHeight)$); 
			\draw [] ($(p3upBL) + (0,\BoardHeightmod)$)
				-- (p3upBL)
				-- ($(p3upBL) + (\BoardWidth,0)$)
				-- ($(p3upBL) + (\BoardWidth, \BoardHeightmod)$); 

			\fill[top color = gray!100,bottom color = gray!25, middle color=gray] (p4lowUL)
				-- ($(p4lowUL) + (\BoardWidth,0)$)
				--  ($(p4lowUL) + (\BoardWidth,-1 * \BoardHeight)$)
				-- ($(p4lowUL) + (0,-1 * \BoardHeight)$)
				-- cycle; 
			\draw [] (p4lowUL)
				edge [densely dotted]  ($(p4lowUL) + (0,-1 * \BoardHeight)$); 
			\draw [] ($(p4lowUL) + (\BoardWidth,0)$)
				edge [densely dotted]  ($(p4lowUL) + (\BoardWidth,-1 * \BoardHeight)$); 
			\draw [] ($(p4lowUL) + (0,-1 * \BoardHeightmod)$)
				-- (p4lowUL)
				-- ($(p4lowUL) + (\BoardWidth,0)$)
				-- ($(p4lowUL) + (\BoardWidth,-1 * \BoardHeightmod)$); 

			\fill[bottom color = gray!100,top color = gray!25, middle color=gray] (p4upBL)
				-- ($(p4upBL) + (\BoardWidth,0)$)
				--  ($(p4upBL) + (\BoardWidth,\BoardHeight)$)
				-- ($(p4upBL) + (0,\BoardHeight)$)
				-- cycle; 
			\draw [] (p4upBL)
				edge [densely dotted]  ($(p4upBL) + (0,\BoardHeight)$); 
			\draw [] ($(p4upBL) + (\BoardWidth,0)$)
				edge [densely dotted]  ($(p4upBL) + (\BoardWidth,\BoardHeight)$); 
			\draw [] ($(p4upBL) + (0,\BoardHeightmod)$)
				-- (p4upBL)
				-- ($(p4upBL) + (\BoardWidth,0)$)
				-- ($(p4upBL) + (\BoardWidth, \BoardHeightmod)$); 

			\fill[top color = gray!100,bottom color = gray!25, middle color=gray] (p5lowUL)
				-- ($(p5lowUL) + (\BoardWidth,0)$)
				--  ($(p5lowUL) + (\BoardWidth,-1 * \BoardHeight)$)
				-- ($(p5lowUL) + (0,-1 * \BoardHeight)$)
				-- cycle; 
			\draw [] (p5lowUL)
				edge [densely dotted]  ($(p5lowUL) + (0,-1 * \BoardHeight)$); 
			\draw [] ($(p5lowUL) + (\BoardWidth,0)$)
				edge [densely dotted]  ($(p5lowUL) + (\BoardWidth,-1 * \BoardHeight)$); 
			\draw [] ($(p5lowUL) + (0,-1 * \BoardHeightmod)$)
				-- (p5lowUL)
				-- ($(p5lowUL) + (\BoardWidth,0)$)
				-- ($(p5lowUL) + (\BoardWidth,-1 * \BoardHeightmod)$); 

			\fill[bottom color = gray!100,top color = gray!25, middle color=gray] (p5upBL)
				-- ($(p5upBL) + (\BoardWidth,0)$)
				--  ($(p5upBL) + (\BoardWidth,\BoardHeight)$)
				-- ($(p5upBL) + (0,\BoardHeight)$)
				-- cycle; 
			\draw [] (p5upBL)
				edge [densely dotted]  ($(p5upBL) + (0,\BoardHeight)$); 
			\draw [] ($(p5upBL) + (\BoardWidth,0)$)
				edge [densely dotted]  ($(p5upBL) + (\BoardWidth,\BoardHeight)$); 
			\draw [] ($(p5upBL) + (0,\BoardHeightmod)$)
				-- (p5upBL)
				-- ($(p5upBL) + (\BoardWidth,0)$)
				-- ($(p5upBL) + (\BoardWidth, \BoardHeightmod)$); 

			\fill[top color = gray!100,bottom color = gray!25, middle color=gray] (p6lowUL)
				-- ($(p6lowUL) + (\BoardWidth,0)$)
				--  ($(p6lowUL) + (\BoardWidth,-1 * \BoardHeight)$)
				-- ($(p6lowUL) + (0,-1 * \BoardHeight)$)
				-- cycle; 
			\draw [] (p6lowUL)
				edge [densely dotted]  ($(p6lowUL) + (0,-1 * \BoardHeight)$); 
			\draw [] ($(p6lowUL) + (\BoardWidth,0)$)
				edge [densely dotted]  ($(p6lowUL) + (\BoardWidth,-1 * \BoardHeight)$); 
			\draw [] ($(p6lowUL) + (0,-1 * \BoardHeightmod)$)
				-- (p6lowUL)
				-- ($(p6lowUL) + (\BoardWidth,0)$)
				-- ($(p6lowUL) + (\BoardWidth,-1 * \BoardHeightmod)$); 

			\fill[bottom color = gray!100,top color = gray!25, middle color=gray] (p6upBL)
				-- ($(p6upBL) + (\BoardWidth,0)$)
				--  ($(p6upBL) + (\BoardWidth,\BoardHeight)$)
				-- ($(p6upBL) + (0,\BoardHeight)$)
				-- cycle; 
			\draw [] (p6upBL)
				edge [densely dotted]  ($(p6upBL) + (0,\BoardHeight)$); 
			\draw [] ($(p6upBL) + (\BoardWidth,0)$)
				edge [densely dotted]  ($(p6upBL) + (\BoardWidth,\BoardHeight)$); 
			\draw [] ($(p6upBL) + (0,\BoardHeightmod)$)
				-- (p6upBL)
				-- ($(p6upBL) + (\BoardWidth,0)$)
				-- ($(p6upBL) + (\BoardWidth, \BoardHeightmod)$); 

			\node[above, align=center] at ($(p0upBL) + (\BoardWidth / 2, \BoardHeight)$) {\large 0};
			\node[above, align=center] at ($(p1upBL) + (\BoardWidth / 2, \BoardHeight)$) {\large 1};
			\node[above, align=center] at ($(p2upBL) + (\BoardWidth / 2, \BoardHeight)$) {\large 2};
			\node[above, align=center] at ($(p3upBL) + (\BoardWidth / 2, \BoardHeight)$) {\large 3};
			\node[above, align=center] at ($(p4upBL) + (\BoardWidth / 2, \BoardHeight)$) {\large 4};
			\node[above, align=center] at ($(p5upBL) + (\BoardWidth / 2, \BoardHeight)$) {\large 5};
			\node[above, align=center] at ($(p6upBL) + (\BoardWidth / 2, \BoardHeight)$) {\large 6};
			
			\draw[decorate, yshift=-4ex]  ($(p0upBL) + (0,1.4 * \BoardHeight)$)
			-- node[above=0.4ex] {\large Reference arm} ($(p2upBL) + (\BoardWidth,1.4 * \BoardHeight)$);
			\draw[decorate, yshift=-4ex]  ($(p4upBL) + (0,1.4 * \BoardHeight)$)
			-- node[above=0.4ex] {\large Reference arm} ($(p6upBL) + (\BoardWidth,1.4 * \BoardHeight)$);
			\node[above=0.4ex, align=center] at ($(p3upBL) + (\BoardWidth / 2,1.4 * \BoardHeight)$) {\large DUT};
			\draw[stealth-stealth]  ($(p0lowUL) + (\BoardWidth / 2,-1.1 * \BoardHeight)$)
			-- node[above] {\large \SI{25}{\milli\metre}} ($(p1lowUL) + (\BoardWidth / 2,-1.1 * \BoardHeight)$);
			\draw[stealth-stealth]  ($(p1lowUL) + (\BoardWidth / 2,-1.1 * \BoardHeight)$)
			-- node[above] {\large \SI{25}{\milli\metre}} ($(p2lowUL) + (\BoardWidth / 2,-1.1 * \BoardHeight)$);
			\draw[stealth-stealth]  ($(p2lowUL) + (\BoardWidth / 2,-1.1 * \BoardHeight)$)
			-- node[above] {\large \SI{58}{\milli\metre}} ($(p3lowUL) + (\BoardWidth / 2,-1.1 * \BoardHeight)$);
			\draw[stealth-stealth]  ($(p3lowUL) + (\BoardWidth / 2,-1.1 * \BoardHeight)$)
			-- node[above] {\large \SI{17}{\milli\metre}} ($(p4lowUL) + (\BoardWidth / 2,-1.1 * \BoardHeight)$);
			\draw[stealth-stealth]  ($(p4lowUL) + (\BoardWidth / 2,-1.1 * \BoardHeight)$)
			-- node[above] {\large \SI{25}{\milli\metre}} ($(p5lowUL) + (\BoardWidth / 2,-1.1 * \BoardHeight)$);
			\draw[stealth-stealth]  ($(p5lowUL) + (\BoardWidth / 2,-1.1 * \BoardHeight)$)
			-- node[above] {\large \SI{23}{\milli\metre}} ($(p6lowUL) + (\BoardWidth / 2,-1.1 * \BoardHeight)$);
			
			\draw[stealth-stealth] ($(p3upBL) + (0,0)$)
				-- node[above] {\large $r$} ($(p3upBL) + (-1 * \DutRadius,0)$);
				
			
			\path  ($(p2upBL) + (\CoordinateDist,-1 * \DutHeight)$)
				pic {vector out};
			\node[right, align=center] at  ($(p2upBL) + (\CoordinateDist + \VecRad,-1 * \DutHeight)$) {\large Column};
			\draw[densely dotted, thick] ($(p2upBL) + (\CoordinateDist,-1 * \DutHeight)$)
				-- ($(p2upBL) + (\CoordinateDist,-1 * \DutHeight + \VecRad)$);
			\draw[-{Latex[width=\ArrWidth, length=\ArrLength]}, thick] ($(p2upBL) + (\CoordinateDist,-1 * \DutHeight + \VecRad)$)
				-- ($(p2upBL) + (\CoordinateDist,0)$);
			\node[right, align=center] at   ($(p2upBL) + (\CoordinateDist + \VecRad,0)$) {\large Row};
			
			\path  ($(p3upBL) + (\CoordinateDist,-1 * \DutHeight / 2)$)
				pic {vector out};
			\node[below, align=center, xshift = 1ex] at  ($(p3upBL) + (\CoordinateDist,-1 * \DutHeight / 2 - \VecRad)$) {\large $x$};
			\draw[densely dotted, thick] ($(p3upBL) + (\CoordinateDist,-1 * \DutHeight / 2)$)
				-- ($(p3upBL) + (\CoordinateDist,-1 * \DutHeight / 2 + \VecRad)$);
			\draw[-{Latex[width=\ArrWidth, length=\ArrLength]}, thick] ($(p3upBL) + (\CoordinateDist,-1 * \DutHeight / 2 + \VecRad)$)
				-- ($(p3upBL) + (\CoordinateDist,-1 * \DutHeight / 2 + \VecRad + \CoordinateLength)$);
			\node[above, align=center, xshift = 1ex] at  ($(p3upBL) + (\CoordinateDist,-1 * \DutHeight / 2 + \VecRad + \CoordinateLength)$) {\large $y$};
			\draw[densely dotted, thick] ($(p3upBL) + (\CoordinateDist,-1 * \DutHeight / 2)$)
				-- ($(p3upBL) + (\CoordinateDist + \VecRad,-1 * \DutHeight / 2)$);
			\draw[-{Latex[width=\ArrWidth, length=\ArrLength]}, thick] ($(p3upBL) + (\CoordinateDist + \VecRad,-1 * \DutHeight / 2)$)
				-- ($(p3upBL) + (\CoordinateDist + \CoordinateLength,-1 * \DutHeight / 2)$);
			\node[above, align=center] at ($(p3upBL) + (\CoordinateDist + \CoordinateLength,-1 * \DutHeight / 2)$) {\large $z$};	
	\end{tikzpicture}
\end{scaletikzpicturetowidth}

%% file: 4_analysis.tex
\section{Analysis tools and methods} 
\label{sec:analysis-tools}

Data were processed in the Corryvreckan test beam reconstruction software framework~\cite{corryvreckan} by fitting straight lines to clusters found in the six reference planes and interpolating the tracks to the DUT. Event and track quality selection criteria were applied to ensure a clean data sample: precisely one track per event, good straightness ($\chi^2/\text{NDF} < 3$) of the track, and track points on each reference plane. Pixels with too large overall firing quantities (more than \num{1000}~times the average) were ignored. On the DUT, clusters with a distance of below~\SI{250}{\um} are matched to the track. The size of the search window was chosen as large as possible without impacting the statistics due to border exclusion regions. Given the one track per event requirement, the implemented trigger logic (Section~\ref{sec:setup}), extremely low ALPIDE noise \cite{ALPIDE-proceedings-1, ALPIDE-proceedings-2, ALPIDE-proceedings-3} and masking of noisy pixels, the association of clusters to non corresponding tracks is considered negligible.

The reference planes and the DUT are aligned to each other by software using a track-cluster residual minimisation, allowing the planes to move in $x$, $y$ and to rotate around $z$. In a second iteration, the DUT alone was allowed to rotate around the other two axes and to move in $z$. In this last step, also the bending radius was used as an optimisation parameter, resulting in a data-driven estimation of this quantity.

As discussed in Section~\ref{sec:bending}, a part of the DUT containing a small fraction of the pixel matrix (\num{\approx30}~rows) is glued to the carrier card. Due to the additional scattering from the carrier card, the sensitivity to the geometrical model used to describe the DUT is reduced in this region. Therefore, the bent shape of the DUT was approximated as a purely cylindrical segment in this analysis, without considering the flat part ($y_0$ in Section~\ref{sec:curvature_measurement}).

The efficiency of the DUT is then estimated by the fraction of tracks with associated clusters. The relative uncertainties are obtained by calculating the Clopper-Pearson interval.

%% file: 5_results.tex
\section{Results}
\label{sec:results}
\subsection{Cluster size, curvature radius and residuals} 
\label{sec:single-run-analysis}

The results presented in this section are based on the data from a single measurement (run) where the DUT was operated at the nominal conditions i.e.~a threshold of \SI{100}{\ele}. To fully illuminate the DUT, the measurement was repeated in two DUT positions with a relative shift along the $y$-axis (row direction, see~Fig.~\ref{fig:setup_sketch}) of \SI{2.5}{\mm}, of which the lower one is shown here.

\begin{figure}[!thp]
	\centering
    \includegraphics[page=1,width=\textwidth]{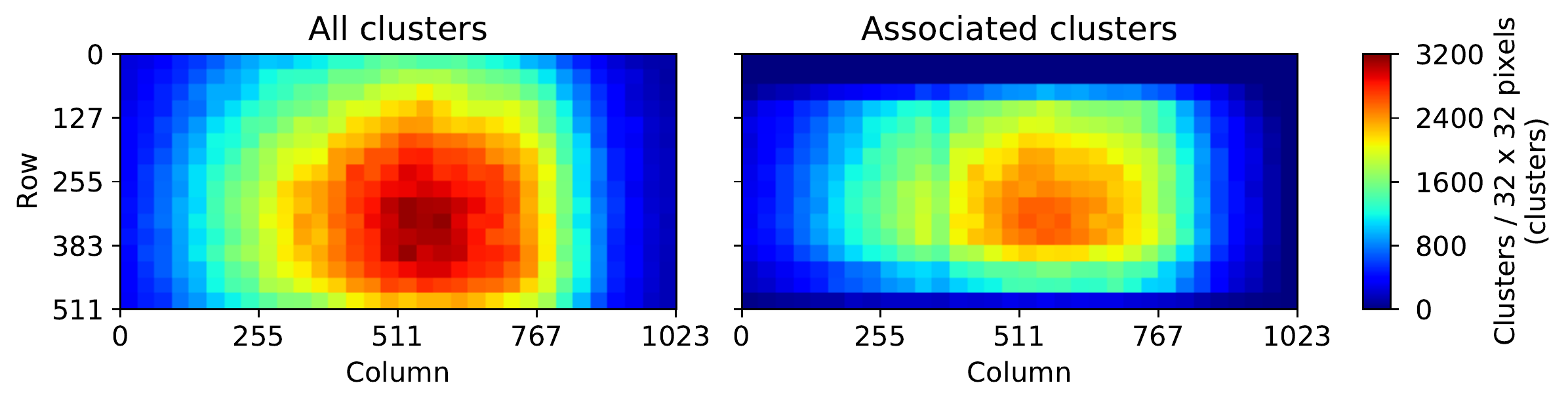}
	\caption{Distributions of all the clusters on the DUT (left) and of those associated to the tracks (right). Fewer clusters are associated to tracks in columns 376-381 due to exclusion of a dead double column.}
	\label{fig:clusters-2D}
\end{figure}

The positions of all the clusters on the DUT and those associated to a track are shown in figure~\ref{fig:clusters-2D}.
In the area near row~512, where the chip is glued to the carrier card, less associated clusters are found given the lower number of reconstructed tracks (Section~\ref{sec:analysis-tools}).
On the opposite side, i.e.~near row 0, there are no clusters associated to tracks due to the relative position of the DUT with respect to the reference planes.
The associated cluster distribution is further affected by the relative position of the reference planes; given the precision of mechanical alignment of the chip position of few millimeters and the requirement of a hit in all reference planes for the reconstructed tracks, the effective area illuminated by the beam is smaller than the chip size. This effect is mostly notable as a drop-off in the number of associated clusters near rows~100 and~400. 

\begin{figure}[!tbhp]
    \centering
    \includegraphics[width=0.7\textwidth]{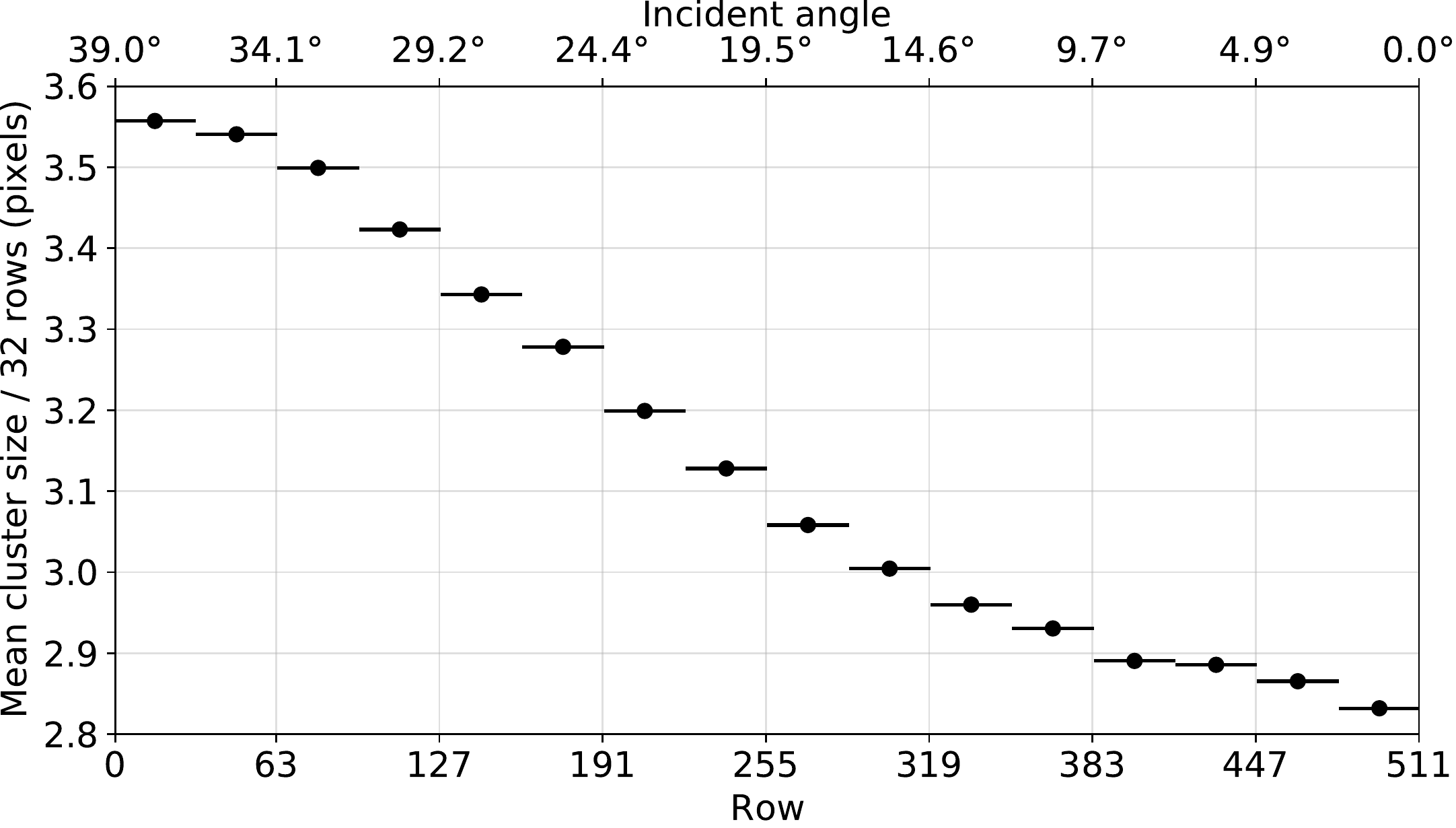}
	\caption{Average cluster size as a function of row and incident angle. The average cluster size decreases with the increasing row number; as the track incident angle decreases, so does the interaction volume and thus the deposited charge.}
	\label{fig:clustersize-vs-row}
\end{figure}

From the data in Fig.~\ref{fig:clusters-2D} (left), the average cluster sizes were calculated for 16 groups of 32 rows and shown in Fig.~\ref{fig:clustersize-vs-row}. With the increasing row number, the incident angle of the beam with respect to the DUT decreases, thus decreasing the particle path in the active volume and therefore the deposited charge, finally resulting in smaller clusters.

The least squares optimisation of the cylindrical model (Section~\ref{sec:analysis-tools}) yielded the DUT radius of \SI[separate-uncertainty = true]{22(1)}{\mm}. The uncertainty on the radius takes into account the variation of the least square optimisation result over all the runs.
The impact of the beam profile on this result was evaluated using two sets of tracks: the first uniformly distributed over the illuminated DUT surface and the second using all tracks with associated clusters on the DUT, as visible in Fig.~\ref{fig:clusters-2D} (right). Both sets result in the same curvature radius.

\begin{figure}[thp]
	\centering
    \includegraphics[width=\textwidth]{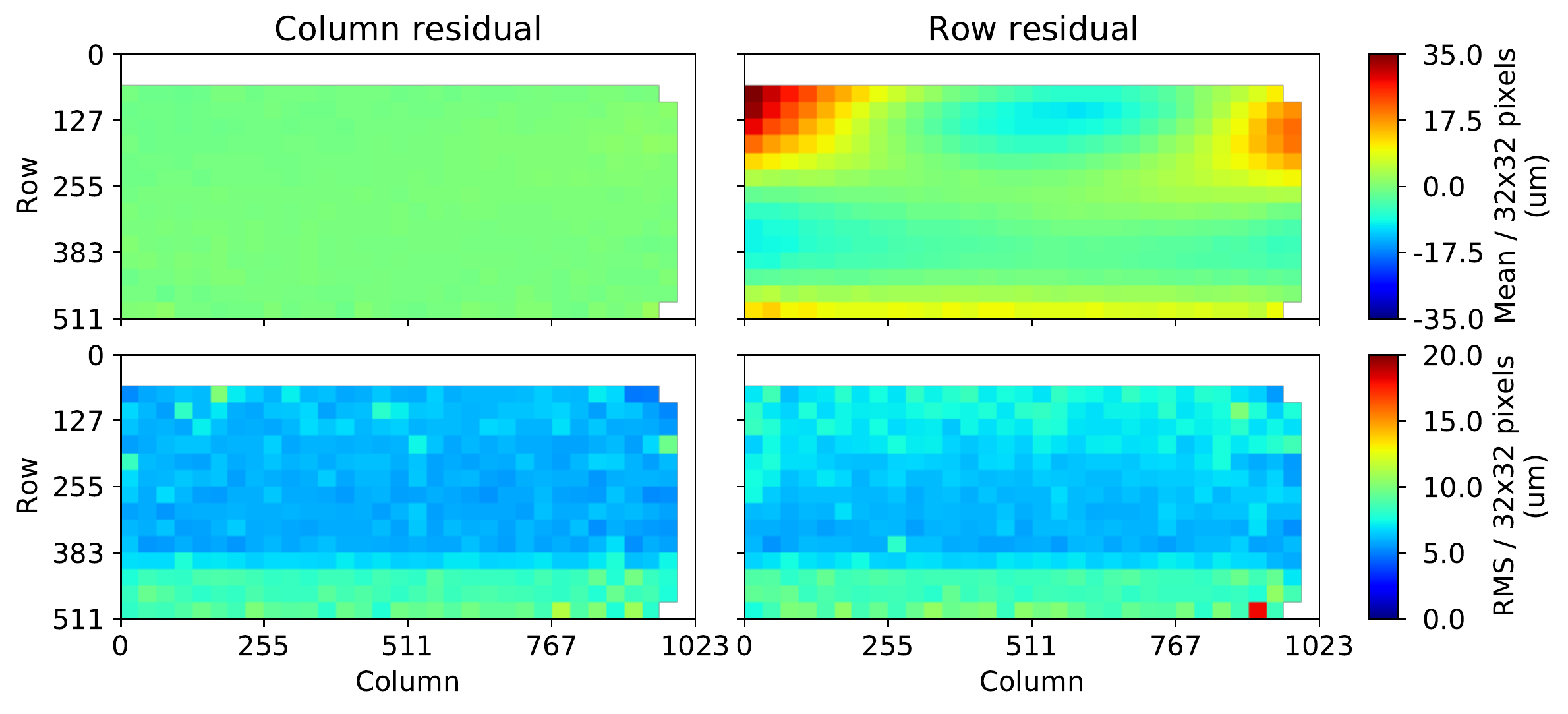}
	\caption{Mean (top panels) and RMS (bottom panels) of the column and row residuals. The cylindrical model description is compatible with the data within \SI{35}{\um}.}
	\label{fig:residual-2D}
\end{figure}

The mean and the RMS of the residuals in the column and row directions are shown in Fig.~\ref{fig:residual-2D}.
A systematic effect of magnitude of up to~\SI{35}{\um} can be observed in the row residual mean, most prominent in the unattached corners of the DUT and along the glued edge.
The RMS of both residuals above row~400 increases, which is compatible with the position of the carrier card, i.e.~can be attributed to the increase of the multiple scattering.
Also an increase of the row residual RMS with  decreasing row number can be observed, a trend compatible with the cluster size increase that is observed for larger beam incident angles (Fig.~\ref{fig:clustersize-vs-row}).

\subsection{Detection efficiency} 
\label{sec:efficiency}

The data from different runs were combined to evaluate the efficiency at different thresholds and over the entire DUT surface. 
The border region equivalent to the track association window size (\SI{250}{\um} i.e.~\num{9}~pixels), as well as the same width region on each side of a dead double-column (columns 369 to 388) were excluded from the efficiency calculation.

\begin{figure}[thp]
    \centering
    \includegraphics[page=1, width=.9\textwidth]{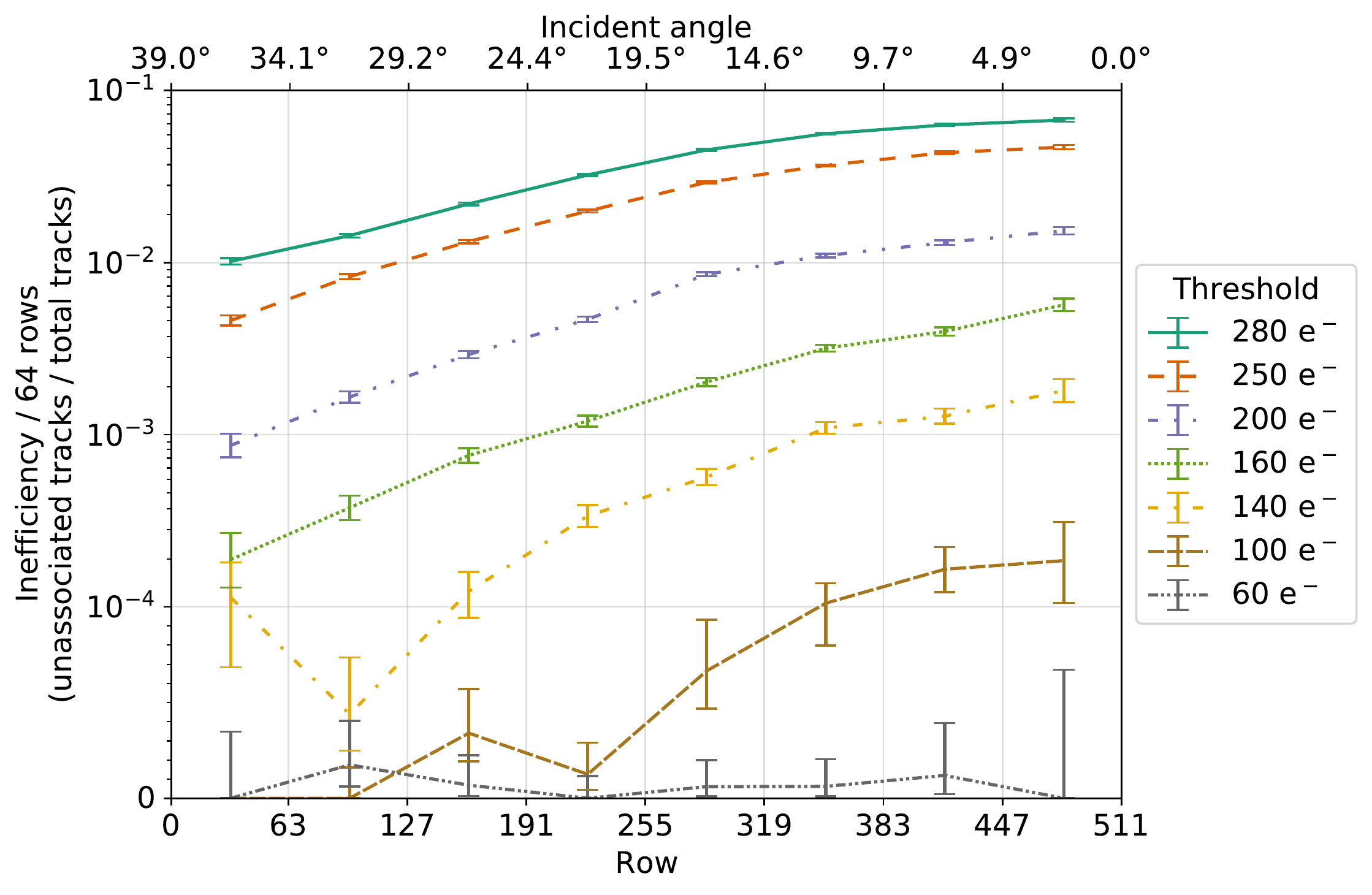}
	\caption{Inefficiency as a function of row and incident angle for different thresholds with partially logarithmic scale ($10^{-1}$ to $10^{-5}$) to show fully efficient rows. The error bars indicate the statistical uncertainty. Each data point corresponds to at least 28k~tracks.}
	\label{fig:efficiency-vs-row}
\end{figure}

\begin{figure}[thp]
    \centering
    \includegraphics[page=1,clip, trim=3.5cm 0cm 21.5cm 1.5cm, width=\textwidth]{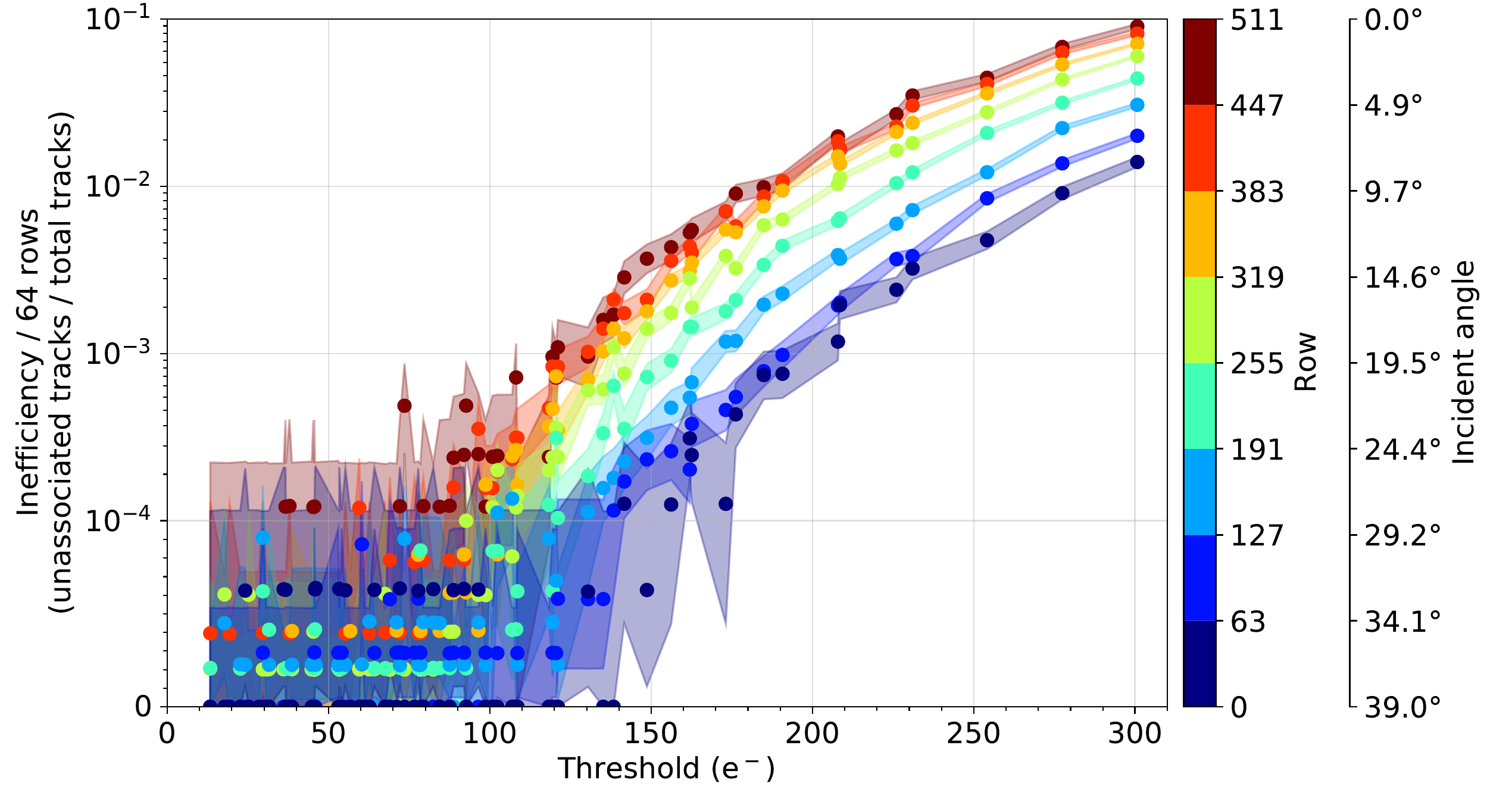}
	\caption{Inefficiency as a function of threshold for different rows and incident angles with partially logarithmic scale ($10^{-1}$ to $10^{-5}$) to show fully efficient rows. The dark circles represent the calculated efficiency and the shaded areas the statistical uncertainty. Each data point corresponds to at least 8k~tracks.}
	\label{fig:efficiency-vs-threshold}
\end{figure}

Figures~\ref{fig:efficiency-vs-row} and~\ref{fig:efficiency-vs-threshold} show the inefficiency as a function of row, beam incident angle, and threshold. Each data point corresponds to at least 8k~tracks, and over 48k~tracks for central rows (given the beam profile, Fig.~\ref{fig:clusters-2D}).
For threshold above \SI{100}{\ele} (the nominal operating point of ALPIDE), the efficiency increases with increasing beam incident angle (decreasing row number). Below~\SI{100}{\ele}, the inefficiency is generally lower than~\num{e-4}, showing that an excellent detecting performance is retained.

%% file: 6_summary.tex
\section{Summary}
The feasibility of bent MAPS was demonstrated for the first time. In particular, \SI{50}{\um}-thick ALPIDE chips were measured in the laboratory and in a beam test while being bent to radii of about \SI{22}{\mm}. They show no sign of any deterioration in operation. Their charge thresholds remain unaffected by the bending and detection efficiencies are measured to largely exceed \SI{99.9}{\percent} without any visible systematic degradation across the full chip surface. 

These very encouraging results do not only mark an important milestone in the R\&D carried out for the ALICE ITS3, but generally open the way to highly integrated, silicon-only, bent sensor arrangements. A new class of detector designs featuring ideal geometries and yielding unprecedented performance figures is at reach.

%% file: acknowledgements.tex
The measurements leading to these results have been performed at the Test Beam Facility at DESY Hamburg (Germany), a member of the Helmholtz Association (HGF). We would like to thank the coordinators at DESY for their valuable support of this testbeam measurement and for the excellent testbeam environment.

We thank the Corryvreckan developers, especially S.~Spannagel and L.~Huth for their precious support.

B.M.~Blidaru acknowledges support by the HighRR research training group [GRK~2058].

F.~Krizek acknowledges support by the Ministry of Education, Youth and Sports of the Czech Republic project [LM2018104].

G.~Usai acknowledges support by the project ``STITCHED MAPS'' funded by MIUR  -- grant PRIN 2017C7KLSX -- Italy

%% file: authors.tex
\section{The ALICE ITS project}\label{sec:authors}
\begin{flushleft} 

G.~Aglieri Rinella$^{\rm 18}$, 
M.~Agnello$^{\rm 14}$, 
B.~Alessandro$^{\rm 27}$, 
F.~Agnese$^{\rm 47}$,
R.S.~Akram$^{\rm 3}$, 
J.~Alme$^{\rm 7}$, 
E.~Anderssen$^{\rm 32}$,
D.~Andreou$^{\rm 33}$,
F.~Antinori$^{\rm 25}$, 
N.~Apadula$^{\rm 32}$
% ...

\bigskip 

% ...
$^{3}$ COMSATS University Islamabad, Islamabad, Pakistan\\
% ...
$^{7}$ Department of Physics and Technology, University of Bergen, Bergen, Norway\\
% ...
$^{14}$ Dipartimento DISAT del Politecnico and Sezione INFN, Turin, Italy\\
% ...
$^{18}$ European Organization for Nuclear Research (CERN), Geneva, Switzerland\\
% ...
$^{25}$ INFN, Sezione di Padova, Padova, Italy\\
% ...
$^{27}$ INFN, Sezione di Torino, Turin, Italy\\
% ...
$^{32}$ Lawrence Berkeley National Laboratory, Berkeley, California, United States\\
$^{33}$ Nikhef, National institute for subatomic physics, Amsterdam, Netherlands\\
% ...
$^{47}$ Universit\'{e} de Strasbourg, CNRS, IPHC UMR 7178, F-67000 Strasbourg, France\\
% ...

\end{flushleft}